\documentclass[journal]{IEEEtran}
\usepackage{cite}
\usepackage{amsmath,amssymb,amsfonts}
\usepackage{algorithm,algorithmic}
\usepackage{graphicx}
\usepackage{textcomp}
\usepackage{wrapfig}
\begin{document}
\title{Interference Aware Cooperative Routing for Edge Computing-enabled 5G Networks}
\author{Abdullah Waqas, Hasan Mahmood, and Nasir Saeed, \IEEEmembership{Senior Member, IEEE}
\thanks{Abdullah Waqas and Nasir Saeed are with the Department of Electrical Engineering, National University of Technology (NUTECH), Islamabad, Pakistan\\
E-mail: abdullah@nutech.edu.pk; mr.nasir.saeed@ieee.org }
\thanks{Hasan Mahmood is with the Department of Electronics, Quaid-I-Azam University, Islamabad, Pakistan\\
Email: hassan@qau.edu.pk}}

\maketitle
\begin{abstract}
Recently, there has been growing research on developing interference-aware routing (IAR) protocols for supporting multiple concurrent transmission in next-generation wireless communication systems. The existing IAR protocols do not consider node cooperation while establishing the routes because motivating the nodes to cooperate and modeling that cooperation is not a trivial task. In addition, the information about the cooperative behavior of a node is not directly visible to neighboring nodes. Therefore, in this paper, we develop a new routing method in which the nodes' cooperation information is utilized to improve the performance of edge computing-enabled 5G networks. The proposed metric is a function of created and received interference in the network. The received interference term ensures that the Signal to Interference plus Noise Ratio (SINR) at the route remains above the threshold value, while the created interference term ensures that those nodes are selected to forward the packet that creates low interference for other nodes. The results show that the proposed solution improves ad hoc networks' performance compared to conventional routing protocols in terms of high network throughput and low outage probability.
\end{abstract}

\begin{IEEEkeywords}
Cooperative Routing, Edge Computing-enabled 5G Networks, Reactive Routing
\end{IEEEkeywords}
\section{Introduction}
\label{sec:introduction}
\IEEEPARstart{T}{he} 3GPP release 16 broadly categorizes the services offered by fifth-generation (5G) networks into three areas, i.e., enhanced Mobile Broadband (eMBB), Ultra-Reliable Low Latency Communication (URLLC), and massive Machine Type Communication (mMTC) \cite{Viet2020}.  Therefore, 5G networks are designed to serve such diverse applications efficiently, utilizing cloud-native concepts like leveraging operations within and across data centers, communicating in a micro-service environment, and simultaneously providing services and applications \cite{Tuyen2017}. A few of the enabling technologies like millimeter wave (mmWave) communication, multiple-input and multiple-output (MIMO), beamforming and small cells are innovations credited to the realization of 5G. In addition, the cloud-native 5G architecture offers a ubiquitous, convenient, and on-demand network access to a shared pool of configurable computing resources (e.g., networks, servers, storage, applications, and services) that can be rapidly provisioned and released with minimal management effort or service provider interaction.

Until fourth-generation (4G), the wireless cellular communication systems focused on increasing the data rates. However, to address the ever-growing need for video streaming, application usage, and massive IoT devices, the cellular network architecture needs to be updated to a more flexible and service-oriented design. Therefore, 5G wireless communications systems have a more distributed architecture than its preceding generations with the introduction of control and user plane separation (CUPS) of the Evolved Packet Core (EPC) \cite{Mamta2021}. For example, the deployment of Unmanned Aerial Systems (UAS) swarms network can yield high capacities to meet the accuracy, stability, and effective service requirements for UAS swarm \cite{wang20205g,wang2021reinforcement,wang2021extensive}. From a mobile network operator (MNO) perspective, the primary motivations for 5G to have a distributed architecture are: to deliver high bandwidth and support a high density of users by utilizing small cells; data processing at the edge of the network, i.e., close to the source helps meet the latency requirements and increasing efficiency; and efficient energy utilization and enhanced data security.

The building block of edge computing-enabled 5G networks  is infrastructure less network of wireless devices, in which, the devices act cooperatively to establish communication between a source and a destination \cite{Guangming2021}. These networks have applications in many real life scenarios such as military communication, rescue and emergency operations, conference rooms, and in the situation where infrastructure is not available or difficult to deploy \cite{ref1}. On the other hand, the distributed nature of these networks imposes multiple challenges such as data privacy, network security, and detection and mitigation of malicious devices\cite{wang2021counter,liu2020deep}. As there is no central management unit, the devices themselves manages network operations. Therefore, each node act as router/relay and forwards the data of other nodes in addition to its own data packet \cite{ref2}. The limitation on the available battery power at each node, requires that the designed algorithm is energy efficient to increase the network life time.

Interference is an important parameter that affects the energy consumption and efficiency of edge computing-enabled 5G networks \cite{Guangming2021}. For example, in networks with Code Division Multiple Access (CDMA), a large bandwidth is used to transmit signals; as a result, the system becomes resilient to noise, jamming, unauthorized interception, and perception. These characteristics exhibit desirable properties, but the system's performance is compromised in the presence of uneven transmission power distribution, especially difficult to control in peer-to-peer networks. Due to independence in the decision-making, there is an inherent imbalance in adjusting power, which in turn results in the near-far-effect \cite{ref3}. To improve the signal to noise plus Interference Ratio (SINR), each node adjusts transmit power, disregarding the levels used by other nodes. As a result, the energy consumed by the network is directly influenced by the amount of interference \cite{kundaliya2020cl}.

There are many methods proposed in the literature to reduce interference among the nodes in various wireless networks \cite{khan2017energy, he2019interference, ahmed2018rp}. Although these schemes select routes with less interference, no mechanism exists that reduces the interference in the network. For example, in \cite{amiri2018interference}, the authors addressed this issue by proposing a solution to reduce interference by the secondary users to the primary user using game theory in cognitive radio networks.  Similarly, the solution in \cite{khan2017energy} reduces interference during routing operation in underwater wireless sensor networks, which selects the next hop in the route having less number of neighbors to improve the SINR of the route. The authors in \cite{chai2020delay} propose a different approach, where the routing algorithm selects the paths with a minimum delay while the delay is computed based on the number of interfering nodes in the vicinity of the source node. Nevertheless, the existing solutions use a selfish approach where each node is concerned about its utility without considering the effects of its actions on the communication between the other nodes of the network. Therefore, in this work, we propose an algorithm to reduce interference in the network in a cooperative way. The route between the nodes is selected in the proposed method such that each node creates minimum interference for the other nodes while ensuring that its own SINR does not decrease below a threshold SINR level. The contributions of the paper are summarized as follows:
\begin{enumerate}
    \item We developed a novel routing algorithm that considers the created interference by a node in the network as a parameter while making the routing decisions.
    \item The proposed algorithm also considers the cooperation parameter in metric, models practical limitations where some nodes may not be willing to cooperate due to their bad channel conditions.
    \item The algorithm enables only those nodes as a relay that creates low interference in the network. Thus, it reduces the energy consumption of the nodes by reducing interference in the network.

\end{enumerate}

The rest of the paper is organized as follows; in Section \ref{literaturereview}, we present a review of energy-efficient routing algorithms. In Section \ref{iacr}, we explain the proposed Interference Aware Cooperative Routing (IACR) Algorithm in detail. In Section \ref{perevl}, the performance of the proposed scheme is discussed for various network parameters, followed by conclusions in Section \ref{conclusion}.

\section{Literature Review}
\label{literaturereview}
In the previous section, we discussed that a substantial portion of the energy resources is consumed during network management by mitigating the interference and packet forwarding operations. While the nodes have limited battery resources, avoiding interference and energy efficiency is an essential requirement to increase the operational time of the network. We can divide the methods to minimize interference in a distributed network into three broader categories: power control, channel scheduling, and routing-based methods. The power control methods aim to find the optimal transmission power range so that the network's interference is minimized. On the other hand, the channel scheduling methods divide the available bandwidth into various sub-bands and allocate the channels to the users such that the interference is reduced. Alternatively, routing-based methods establish the path between the source and destination such that overall interference in the network is minimized.

There are various power control techniques for minimizing network interference. For instance, in \cite{ji2020power}, the authors propose a power control algorithm for cognitive vehicular networks by obtaining the optimal power vector considering the minimum value of outage probability of the system. The proposed system performs better than other power allocation techniques. However, the reduced transmission power increases the number of hops in the route; as a result, the algorithm introduces transmission delays during communication. In \cite{rodrigues2019edge}, the authors apply Particle Swarm Optimization (PSO) technique to reduce transmission delays by updating transmission powers for the nodes in each iteration. Although the proposed scheme reduces time delays, on the other hand, the assignment of different power levels to each node in the network may be counterproductive in terms of reduced network connectivity and variable interference in the network. In \cite{chincoli2016power}, the authors analyzed the performance of the distributed sensor network in the presence of variable interference. The study compares multiple power control methods and concludes that using homogeneous transmission powers for all the nodes in the network results in better network performance in the scenarios where interference is variable. However, finding the minimum transmission power to ensure connectivity is not straightforward, especially in the case of homogeneous power assignment. Moreover,  the effect of step size on the performance of the algorithm needs to be evaluated.

Unlike the power control techniques, in \cite{hisham2020adjacent}, the authors formulate joint scheduling and power control problem as Mixed Boolean Linear Programming (MBLP) problem in order to mitigate the impact of co-channel and adjacent channel interference on vehicle-to-vehicle communication. The results of \cite{hisham2020adjacent} show that the impact of channel interference can be reduced by scheduling and power control. However, the solution obtained is numerically sensitive, where a stable optimal solution can be obtained by applying sensitivity reduction techniques at the cost of increased computational complexity.

Consequently, in \cite{unlu2019ipbm}, the authors proposed a solution to handle interference during the communication of control packets, dividing one frame into three types of slots where data is sent on each slot by a fair scheduling algorithm. Thus, the rescheduling is performed based on the interference pattern at each time slot. The proposed algorithm in \cite{unlu2019ipbm} can be integrated with distributed routing protocols such as AODV and ALOHA with fixed frame lengths. However, the performance of the solution needs to be analyzed for variable frame structure and dynamic interference conditions.

The third type of algorithms reduce interference at network layer by reducing interference during routing operations. Recently, in \cite{wang2020beamforming}, the authors presented Optimized Ad-hoc On-Demand Distance Vector (OAODV) routing protocol, enhancing the routing and scheduling performance of Unmammded Aerial Systems (UAS).  In \cite{wang2020beamforming}, beam-forming is used to send the data towards the desired destination with  reduced interference. On the other hand, the authors in \cite{waqas2020interference} minimizes the interference during route discovery operation. In the proposed solution, only those node whose interference is less than a predefined threshold value actively participate during routing operations, reducing the energy consumption and achieving high network throughput. While the algorithm ensures that the SINR does not decrease the threshold SINR value, the resulting route may choose nodes at the network's edges, resulting in higher time delays. Similarly, in \cite{waqas2012energy}, the authors defined a routing metric that uses link interference in the network to select the path whose SINR is high to improve the performance of the network. The proposed solution faces a similar problem, i.e.,  high link length results in high time latency during the routing operation.

The solutions mentioned above select low interference paths while establishing routes between the nodes, they do not provide the solution to reduce interference provided by a node to other nodes during route establishment. Therefore, we propose a novel solution based on a new routing metric that considers created interference by a node in the network as a parameter during routing decisions. Moreover, we also use the cooperation parameter that models practical limitations of the network where some nodes may not be willing to cooperate due to bad channel conditions. Our proposed solution enables only those nodes as a relay that creates low interference in the network, further reducing the network's energy consumption. In the next section, we present the proposed interference aware cooperative routing (IACR) algorithm that establishes routes such that the interference provided by the source node to the other nodes in the network is minimized under the constraints that its own SINR does not decrease below the minimum SINR level that is required for successful reception of the data packets.

\section{Interference Aware Cooperative Routing}\label{iacr}
This section proposes the IACR algorithm that reduces interference and maximizes the network's overall performance. The IACR defines the route's cost as a function of created and received interference by a node in the network. The metric value is calculated as the weighted sum of received and created interference. The receive interference term in the proposed routing metric ensures that the data is routed through a low interference path, achieving high SINR at the route. On the other hand, the created interference term selects that path for data transmission that offers less interference in the network. As a result of this strategy, the quality of service (QoS) for all the users in the network improves. In the following, we present the system model and working principle of the proposed solution in detail.
\begin{figure}
\centering
\includegraphics[width=1\columnwidth]{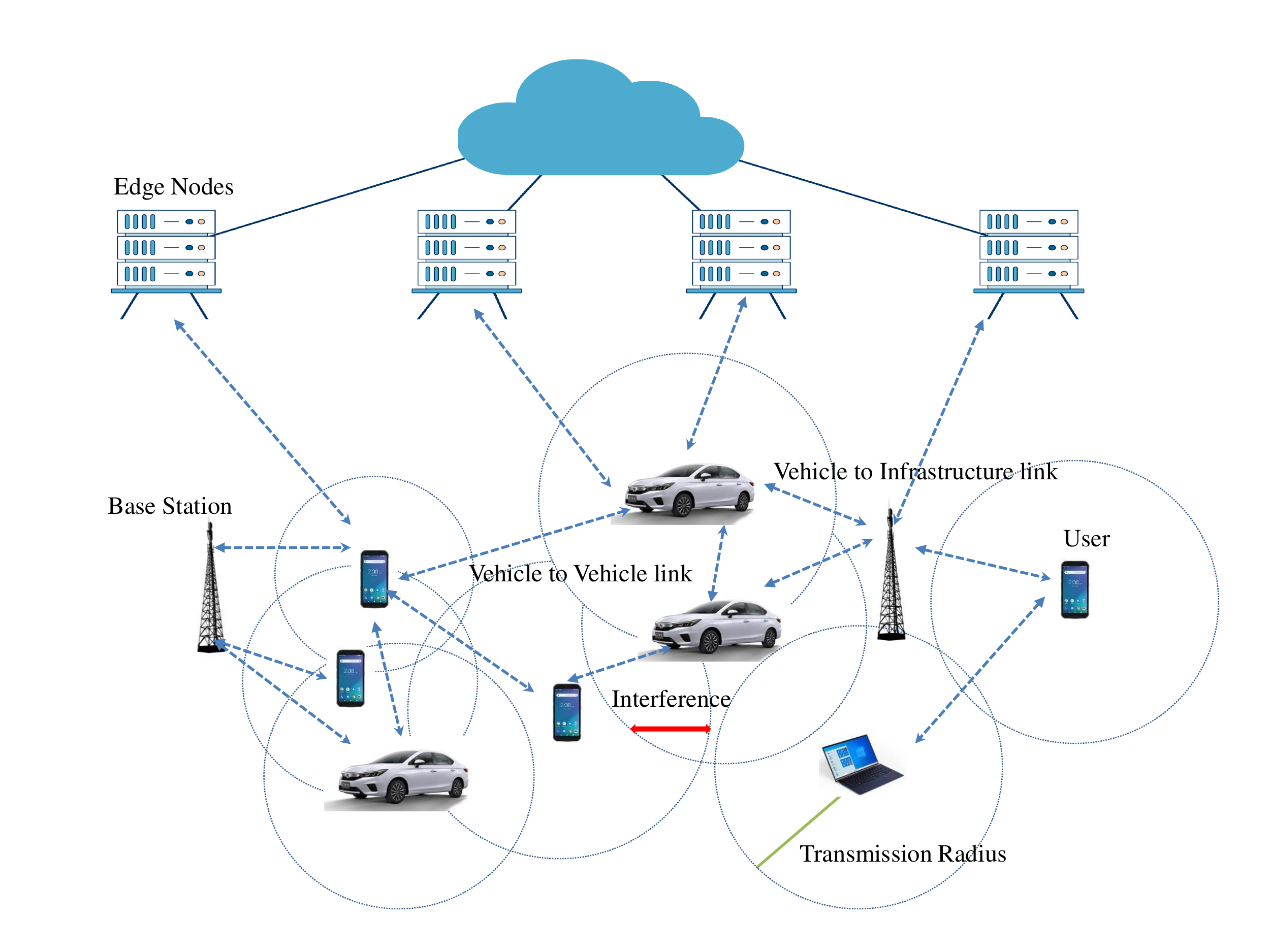} 
\caption{System model}
\label{systemmodel}
\end{figure}
\subsection{System model}
We consider an edge computing-enabled 5G network setup with $N$ number of nodes deployed randomly as shown in Fig. \ref{systemmodel}.  We assume omnidirectional coverage of the nodes to transmit the data in a circular region centered around the transmitting node. The radius of transmission circle $r_t$ depends on the transmission power $P_t$ of the node. The node transmits over the same frequency band. Initially, the nodes start transmission using maximum transmission power $P_{max}$ and update their transmission levels by the proposed method when the communication begins. Then, the received power can be calculated as \cite{ren2013interference, bastos2018assisted} 

\begin{equation}\label{Eq6.1}
  P_r=P_tL_{i,j}=\frac{P_t}{d^\alpha},
\end{equation}
where $\alpha$ is the path-loss exponent and $d$ is the distance between transmitter and receiver.

Additionally, we assume that each node can act as a router for a distributed edge network to configure itself as a source and/or destination. If the destination node is not within the communication range of a source node, multihop routes are used to establish the communication path. Finally, we assume that all nodes are cooperative and do not drop other nodes' packets to save their energy resources.

Since all nodes are operating at the same frequency band, they create interference for each other. We can write the aggregate interference at node $j$ as,
\begin{equation}\label{Eq6.2}
  I_j=\sum\limits_{k=1,k\neq i,j}^{N}{\frac{P_k}{d_{j,k}^\alpha}}
\end{equation}
where $P_k$ is the transmission power of node $k$, $d_{j,k}$ is distance between node $j$ and node $k$, and $N$ is the total number of active nodes in the network. The resultant SINR at node $j$ when receiving data from node $i$ can be calculated as,
\begin{equation}\label{Eq6.3}
  SINR_{(i,j)}=\frac{P_{i,j}}{I_j+\sigma^2}
\end{equation}
where $\sigma^2$ is the noise variance, $I_j$ is received interference at the destination node $j$, and $P_{i,j}$ is power received at node $j$ from the source node $i$. In general, noise power is very less as compared to signal power, therefore, the above equation is reduced as,
\begin{equation}\label{Eq6.4}
  SIR(i,j)=\frac{P_{i,j}}{I_j}
\end{equation}
A packet is successfully received at the destination node if SINR at each link is greater than a predefined threshold level $SINR_{th}$. In the next section, we discuss the proposed IACR algorithm in detail to establish routes between source and destination.
\subsection{Working Principle of IACR algorithm}
The IACR algorithm establishes routes in a distributed fashion similar to the standard ad hoc on-demand distance vector (AODV) routing protocol. However, it uses a new metric field in which the route's cost depends on the interference at the nodes. When the source node desires to establish a communication link, it broadcasts a route request (RREQ) in the network. When each neighbor receives the RREQ packet, it updates the metric value in the header of the RREQ packet and forwards the packet to the next neighbors. The routing metric is updated as,
\begin{equation}\label{Eq6.5}
M_k = M_{RREQ}+M_{k-1,k},
\end{equation}
where $M_k$ is the cost of the route up to $k$-th node, $M_{RREQ}$ is the metric value in the RREQ packet when node $k$ has received the RREQ packet, and $M_{k-1,k}$ is the cost of link $(k-1,k)$. When the packet reaches the destination node, it responds with a route reply (RREP) packet to establish a link with the source node. If the destination node receives multiple RREQ packets, it sends the RREP packet on the route with the minimum value of the routing metric. The routing metric used in the IACR algorithm depends on created and received interference by the node in the network. Each node can calculate the value of received interference using the received power information from the physical layer. However, it can not directly calculate the amount of interference created by itself to other nodes in the network. To get the information about created interference, it transmits an information collection packet towards its neighbors. The information collection phase is explained in detail in the next section. The metric value of a link is calculated as a weighted sum of received and created interference at the node that can be written as,
\begin{equation}
\label{Eq6.6}
M_{i,j}=\delta I_c^i+(1-\delta)I_j,
\end{equation}
where $I_j$ is the interference received at the destination node $j$ that is calculated using (\ref{Eq6.2}), and $\delta$ is the cooperation parameter of the node. The term $I_c^i$ in (\ref{Eq6.6}) represents the interference created by transmitter $i$ to other nodes that can be written as,
\begin{equation}\label{Eq6.7}
  I_c^i=\sum\limits_{k=1,k\neq i}^{N}{\frac{P_{i}}{d(i,k)^\alpha}}
\end{equation}
where $P_{i}$ is transmission power of node $i$ and $d(i,k)$ is the distance between node $k$ and node $i$. The expression in (\ref{Eq6.7}) shows that the interference created by a node depends on its transmission power and its geographical location in the network. The proposed solution works in two phases. In the first phase, it collects necessary information from other nodes to calculate the routing metric's value. The second phase establishes the route towards the destination by selecting those nodes as a relay with a minimum value of routing metric.

\subsubsection{Information Collection Phase}
\begin{table}
\caption{Information table of node $i$}
\centering
\begin{tabular} {|c |c |c |c|}
\hline\hline
Address of the neighbor & $I_c^j$ & $I_j$ & $I_{\text{aggr}}^{j}$\\
\hline
1 & $\frac{P_i}{d(i,1)^\alpha}$ & $\sum\limits_{k=1,k\neq i}^{N}{\frac{P_k}{d(1,k)^\alpha}}$ & $\sum\limits_{k=1,k\neq 1}^{N}{I_c^k}$\\
\hline
2 & $\frac{P_i}{d(i,2)^\alpha}$ & $\sum\limits_{k=1,k\neq i}^{N}{\frac{P_k}{d(2,k)^\alpha}}$ & $\sum\limits_{k=1,k\neq 2}^{N}{I_c^k}$\\
\hline
\vdots & \vdots & \vdots & \vdots \\
\hline
j & $\frac{P_i}{d(i,j)^\alpha}$ & $\sum\limits_{k=1,k\neq i}^{N}{\frac{P_k}{d(j,k)^\alpha}}$ & $\sum\limits_{k=1,k\neq j}^{N}{I_c^k}$\\
\hline
\end{tabular}
\label{table6.1}
\end{table}
As discussed above, the IACR protocol requires information about created and received interference by a node to establish the routes between the source and the destination. The amount of interference created by a node depends on its neighbors' transmission power and geographical position. While a node does not know the location of other nodes in the network, this information is collected by transmitting an information collection packet (ICP) in the network. Algorithm 1 explains the actions of the source and receiving nodes during the information collection phase. As described in Algorithm 1, during the information collection phase, each node $i$ transmits ICP towards its neighbors using maximum transmission power $P_{max}$. Each neighbor $j$ when receives this packet calculates the value of received power ($P_r$) from the node $i$. This value reflects the interference created at node $j$ from node $i$.
Additionally, the receiving node $j$ calculates the value of interference received from other nodes while receiving ICP from node $i$. The node $j$ extracts this information directly from the physical layer while receiving ICP from node $i$. This value reflects the amount of received interference at the link ($i,j$). After calculating $P_r$ and $I_r$, the node $j$ shares this information with node $i$ by sending an information reply packet. The node $i$ repeats this process for all neighbors and saves this information in an information table used in the route establishment phase to take optimal routing decisions.\par
The format of the information table of node $i$ is shown in Table \ref{table6.1}. The first column represents the neighbor's address, and the second column represents the amount of interference created at $j$-th neighbor from node $i$ ($I_c^j$ ). The third column shows the amount of interference experienced by node $j$ while receiving data from node $i$ ($I_j$). The last column represents the aggregate amount of interference created by the node $i$ at other nodes if it selects neighbor $j$ as a relay node in the route ($I_{\text{aggr}}^{j}$ ). This information is used during the route establishment phase to make optimal routing decisions. The transmission of ICP is limited to neighbors of a node only, this information is piggy bagged with “HELLO” and “ACK” packets and does not introduce overhead during routing operation.\par

\begin{algorithm}
 \caption{Information collection phase}
 \begin{algorithmic}[1]
  \STATE N = total number of nodes in the network
  \FOR {node = 1 to $N$}
	\WHILE {acting as transmitter}
        \STATE ICReq = information collection Request packet\;
        \STATE i = number of neighbors of current node\;
        \STATE neighAddr = address of neighbour $i$\;
        \WHILE {$i$ $>$ 0}
            \STATE Destination ID field of ICReq = ID of node $i$\;
            \STATE Transmitter ID field of ICReq = ID of node $i$\;
            \STATE Transmit power field of ICReq = Power of node $i$\;
            \STATE forward ICREQ packet to neighbors\;
            \STATE i = i - 1\;
            \STATE save address of neighbour $i$\;
        \ENDWHILE
    \ENDWHILE
  \WHILE{acting as receiver}
        \IF{received ICreq packet}
            \IF{Destination ID field of ICReq == ID of node}
                \STATE set receive power field of IC reply packet  as receive power $P_r$\;
                \STATE set receive interference field of IC reply packet as interference $I_r$ from neighbors\;
                \STATE set destination ID of IC reply packet as ID of node\;
                \STATE transmit IC reply packet toward source node\;
            \ELSE
                \STATE set receive power field of IC reply packet as receive power $P_r$\;
                \STATE set destination ID of IC reply packet as ID of node\;
                \STATE transmit IC reply packet toward source node\;
            \ENDIF
        \ENDIF
        \IF{IC reply packet received}
            \IF{Transmitter ID field of IC reply packet == ID of node}
                \IF{Destination ID field of IC reply packet contain the address of neighbour}
                    \STATE save signal strangth from the receive power filed of IC reply packet\;
                    \STATE save receive interference value in information table from IC reply packet\;
                \ELSE
                    \STATE save created interference to neighbour in the information table from IC reply packet\;
                \ENDIF
            \ELSE
                \STATE discard the packet\;
            \ENDIF
        \ENDIF
    \ENDWHILE
\ENDFOR
 \end{algorithmic} 
 \end{algorithm}

\subsubsection{Route Establishment Phase}
The IACR protocol establishes routes in a distributed fashion similar to the standard AODV routing protocol. However, in contrast to the AODV protocol, the proposed solution uses interference at the nodes as a routing metric to calculate the cost of a link. The cost of a route is the sum of the costs of individual links. To establish the route at minimum cost, the source node selects those neighbors as potential relays, which minimizes the cost function given in (\ref{Eq6.6}). When the neighboring node receives the RREQ packet during the link establishing phase, it searches its routing table for a route towards the destination node. If new routing information is found, it generates an RREP and sends it towards the source node on the reverse path. In case of unavailability of new routing information, the relay node updates the metric value using (\ref{Eq6.5}) and forwards the packet towards its neighbors. To keep track of the reverse path, the relay node saves the identity of the transmitting node in the routing table. When the RREQ packet reaches the destination node, it acknowledges with an RREP packet on the reverse path. If the destination receives the RREQ packet through multiple paths, it sends RREP on the path with the minimum value of the routing metric. When receiving the RREP packet, each relay node updates the routing table and forwards the packet to the node on the reverse path. When RREP reaches the source, a route is established between source and destination. Algorithm \ref{algo6.2} shows the pseudocode of the route establishment phase for the proposed IACR protocol.\par

\begin{algorithm}
\caption{Establishment of Route}
\label{algo6.2}
\begin{algorithmic}[1]
\IF{source node}
\FOR{all neighbors $k$ of source node $i$}
\STATE calculate metric value $M(k)$ by (\ref{Eq6.6})\;
\ENDFOR
\STATE save $\min(M)$ in RREQ\;
\STATE save ID of node $k$ in RREQ\;
\STATE save destination ID in RREQ\;
\STATE set RREQ transmitter ID as ID of node $k$\;
\STATE set next hop in RREQ as ID of node correspond to $min(M)$\;
\STATE forward RREQ to neighbors\;
        \IF{rrep packet received}
             \STATE update routing table entry with next hop field in RREP\;
             \STATE start data transmission towards next hop\;
        \ENDIF
\ENDIF
\IF{relay node}
\IF{RREQ packet received}
\FOR{all neighbors $k$ of relay node $j$}
    \STATE determine value $M(k)$ by (\ref{Eq6.6})\;
\ENDFOR
              \STATE set RREQ metric value as $\rightarrow$ Current metric+min(M)\;
              \STATE save transmitter ID in reverse path routing table\;
              \STATE set transmitter ID of route request  as ID of node $k$\;
              \STATE update next hop ID as $\rightarrow$ ID of node correspond to min(M)\;
              \STATE forward RREQ to neighbors\;
\ENDIF

\IF{RREP packet received}
            \STATE save transmitter ID in forward path routing table\;
            \STATE set RREP transmitter ID as ID of node $k$\;
\ENDIF
\ENDIF
\IF{destination node}
\IF{RREQ packet received}
    \STATE set source ID in RREP as $\rightarrow$ source ID in RREQ\;
    \STATE set RREP transmitter ID as $\rightarrow$ ID of node $k$\;
    \STATE set next hop in RREP as $\rightarrow$ transmitter ID in RREQ\;
    \STATE set metric of RREP as $\rightarrow$ metric value n RREQ\;
    \STATE forward packet towards source node\;
\ENDIF
\ENDIF
\end{algorithmic}
\end{algorithm}
\section{Performance evaluation}\label{perevl}
In this section, we evaluate the performance of the proposed schemes regarding various network parameters. First, we discuss the simulation setup deployed to assess the performance of the proposed algorithm. Then, we compare the results of IACR with minimum hop count (MHC) and interference-aware energy efficient (IAEE) routing algorithms.
\subsection{Simulation Setup}
The network consists of $N$ wireless devices uniformly distributed in a square area of $A (m^2)$. The nodes work in an ad hoc fashion; therefore, they can also act as routers to forward the packets of their neighboring nodes. We consider that all nodes can transmit equally in all directions with a communication circle of radius $r_t$. The network establishment time is set to 3 seconds, in which a node discovers its neighbors by transmitting ``HELLO” packets in their communication circle. The nodes transmit ``HELLO” packet after every 0.2 seconds to keep an updated routing metric for their neighbors.

When a source node desires to establish a communication link, it initiates a route discovery algorithm and searches for a new route towards the destination before sending data packets. In the simulation setup, multiple nodes are allowed to transmit their data simultaneously. A source node can initiate a request for a link towards the destination at any time after the establishment of the network. When the route is established, the source node transmits a message packet of 512B after every 0.1s towards the destination. The SINR at the receiver is affected by the path-loss factor, and the packet is successfully received at the destination if SINR at the receiver is greater than a predefined threshold value of SINR, which depends on underlying modulation and coding technique. We use binary phase-shift keying (BPSK) modulation in the simulation setup to highlight the effects of the routing on the network's performance.
\subsection{Results and Discussion}
This section discusses the performance of the proposed solution for various network parameters and compares it with a conventional minimum hop count and energy-efficient routing algorithms. In particular, we provide results for the throughput and outage probability of different schemes, which reflect the packet delivery rate of the route. The throughput on a route is defined as the ratio of number of packets successfully received at the destination to the number of packets transmitted by the source node, which can be written as,
\begin{equation}\label{eq_th}
    \hat{\tau} = \frac{n_s}{T_p}
\end{equation}
where $n_s$ is the number of packets successfully received and $T_p$ is the total number of packets transmitted by source node. Then, the outage probability is defined as,
\begin{equation}\label{eq_out}
    P_{out} = Pr[{\gamma \leq \gamma_{th}}]
\end{equation}
where $\gamma$ is the SINR of the route and $\gamma_{th}$ is the threshold value of SINR required for correct reception of the data.

\begin{figure}
\centering
\includegraphics[width=1\columnwidth]{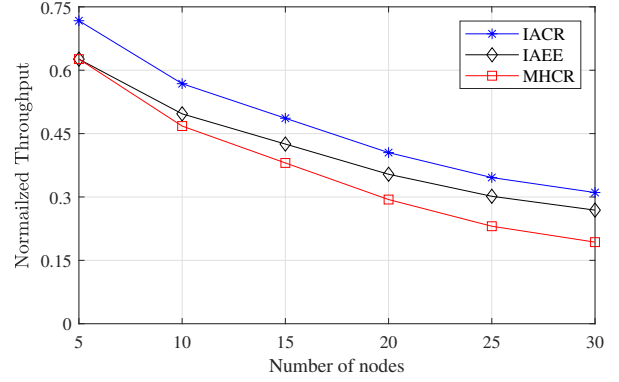} 
\caption{Throughput comparison of IACR, MHC, and IAEE algorithms}
\label{Section6_3_2Figure_1_ThroughputVsNodeDensity}
\end{figure}

In Figure~\ref{Section6_3_2Figure_1_ThroughputVsNodeDensity}, we plot the normalized throughput $\hat{\tau}$ with respect to number of nodes in the network. The graph shows that the value of $\hat{\tau}$ decreases when more nodes initialize their communication. When new nodes take part in communication, the interference in the network increases, which results in decreased network throughput. The graph shows that establishing interference-aware routes between the source and destination improves the throughput of the network. The plot shows that IAEE routing provides higher throughput than the MHC routing algorithm by establishing interference-aware routes between the source and the destination. Still, the proposed solution performs better than IAEE by removing those nodes from the route that provide high interference to other nodes.\par

\begin{figure}
\centering
\includegraphics[width=1\columnwidth]{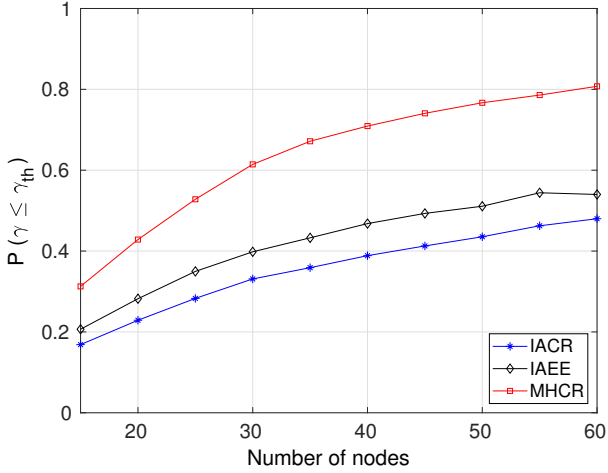} 
\caption{Outage probability of IACR, MHC, and IAEE algorithms}
\label{Section6_3_2Figure_3_outageComparisonNodeDensity}
\end{figure}

Figure~\ref{Section6_3_2Figure_3_outageComparisonNodeDensity} shows the outage probability of the network concerning the number of transmitting nodes in the network. The graph shows that the proposed solution establishes high SINR routes between the source and the destination and reduces the outage probability calculated from (\ref{eq_out}). The graph further illustrates that the outage probability of the route increases when new nodes join the network. When new nodes join the network, the interference in the network increase which consequently reduces the term $\gamma$ in (\ref{eq_out}), and increases in outage probability. The proposed solution adaptive reestablishes the routes when interference at node rises above a threshold value, providing less outage probability than MHCR and IAEE routing algorithms.\par
\begin{figure}
\centering
\includegraphics[width=1\columnwidth]{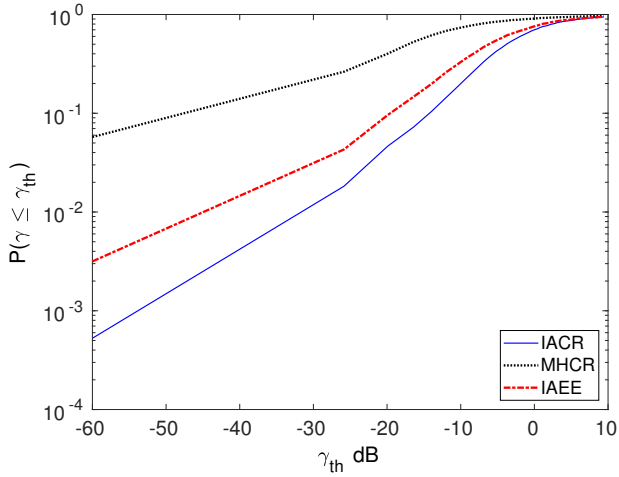} 
\caption{Effects of SINR threshold on outage probability of IACR, IAEE, and MHC algorithms}
\label{Section6_3_2Figure_4_outageComparisonThresholdSINR}
\end{figure}

Figure~\ref{Section6_3_2Figure_4_outageComparisonThresholdSINR} shows the comparison of outage probability of different routing schemes. The results show that the outage probability increases concerning the SINR threshold. At a high SINR threshold, a comparatively high value of received power is required to decode the packet successfully, which results in an increased outage probability. The graph shows that the presented IACR approach performs better than previous schemes, but the performance of IACR is almost equal to the IAEE algorithm when the threshold SINR is high. The graph shows that the outage probability is independent of the choice of the algorithm when the threshold SINR increases to 10 dB. At high threshold values, the performance of the route is dominated by the path-loss factor; therefore, even the SINR at minimum interference routes can not exceed the threshold level; as a result, a high outage is observed. \par
\begin{figure}
\centering
\includegraphics[width=1\columnwidth]{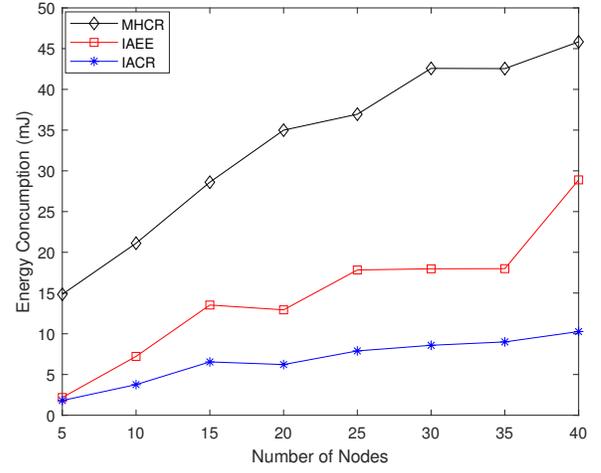} 
\caption{Average energy consumption of network for IACR, IAEE, and MHC algorithms}
\label{Section6_3_2Figure_5_EnergyComparisonNodeDensity}
\end{figure}

Figure \ref{Section6_3_2Figure_5_EnergyComparisonNodeDensity} shows the energy consumption of IACR as compared to previously proposed techniques. The graph shows that the energy consumption of the network increases when the number of nodes increases. Nevertheless, the energy consumption of IACR is low compared to MHCR and IAEE routing protocols because, in IACR, the nodes cooperate and create low interference for each other, which reduces overall interference in the network. When the interference in the network decreases, the nodes can transmit at low transmission power, which saves their energy consumption.

\section{Conclusion and Future Works}\label{conclusion}
In this paper, we present an interference-aware cooperative routing (IACR) algorithm to establish the routes between source and destination in edge computing-enabled 5G networks. In IACR, the nodes cooperate and select those routes for data transmission that minimizes the interference to other nodes. This cooperative behavior reduces overall interference in the network. As a result, the nodes located in dense network regions can achieve better SINR, which improves the network's performance. We show that the proposed IACR routing metric improves network throughput and provides less outage than MHC and IAEE routing algorithms. Furthermore, the performance of the proposed solution improves when the number of nodes in the network is high. Therefore, the proposed schemes can be an efficient interference-aware routing solution for dense edge computing-enabled 5G networks. In the future, we will apply and analyze the performance of the proposed IACR algorithm to large-scale networks. Furthermore, the performance of the proposed algorithm can be tested for various mobility models and dynamic channel conditions covering 5G and beyond networks.

\bibliographystyle{IEEEtran}
\bibliography{references}

\begin{IEEEbiography}[{\includegraphics[width=1in,height=1.25in,clip,keepaspectratio]{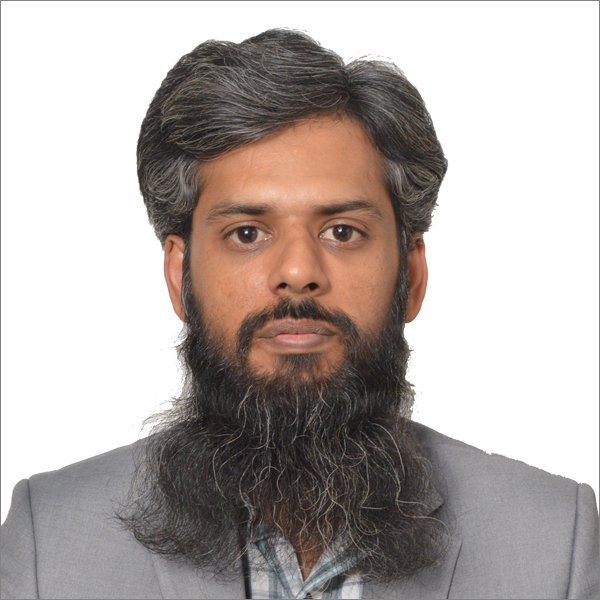}}]{Abdullah Waqas} received  his  M.Phil and  Ph.D.  degrees  from  Quaid-IAzam University, Islamabad, Pakistan in 2010. Currently, he is Assistant Professor in Department of Electrical Engineering, National University of Technology, Pakistan. He worked as faculty member in Department of Electronics, Quaid-I-Azam University, Islamabad, from 2010  to 2015.  His  research  interests  includes  next generation wireless technologies, routing strategies in 5G and 6G networks, network protocols, game theory, and ad hoc and sensor networks.
\end{IEEEbiography}

\begin{IEEEbiography}[{\includegraphics[width=1in,height=1.25in,clip,keepaspectratio]{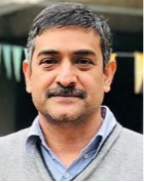}}]{Hasan Mahmood} received the M.Sc. degree in electronics from Quaid-i-Azam University, Islamabad, Pakistan, in 1991, the M.S. degree in electrical engineering from the University of Ulm, Germany, in 2002, and the Ph.D. degree in electrical engineering from the Stevens Institute of Technology, Hoboken, NJ, USA, in 2007. From 1994 to 2000, he was a Faculty Member with the Department of Electronics, Quaid-i-Azam University, where he is currently serving as a Professor. His field of interest are wireless communications, game theory, cryptography, and channel and source coding for wireless systems.
\end{IEEEbiography}

\begin{IEEEbiography}[{\includegraphics[width=1in,height=1.25in,clip,keepaspectratio]{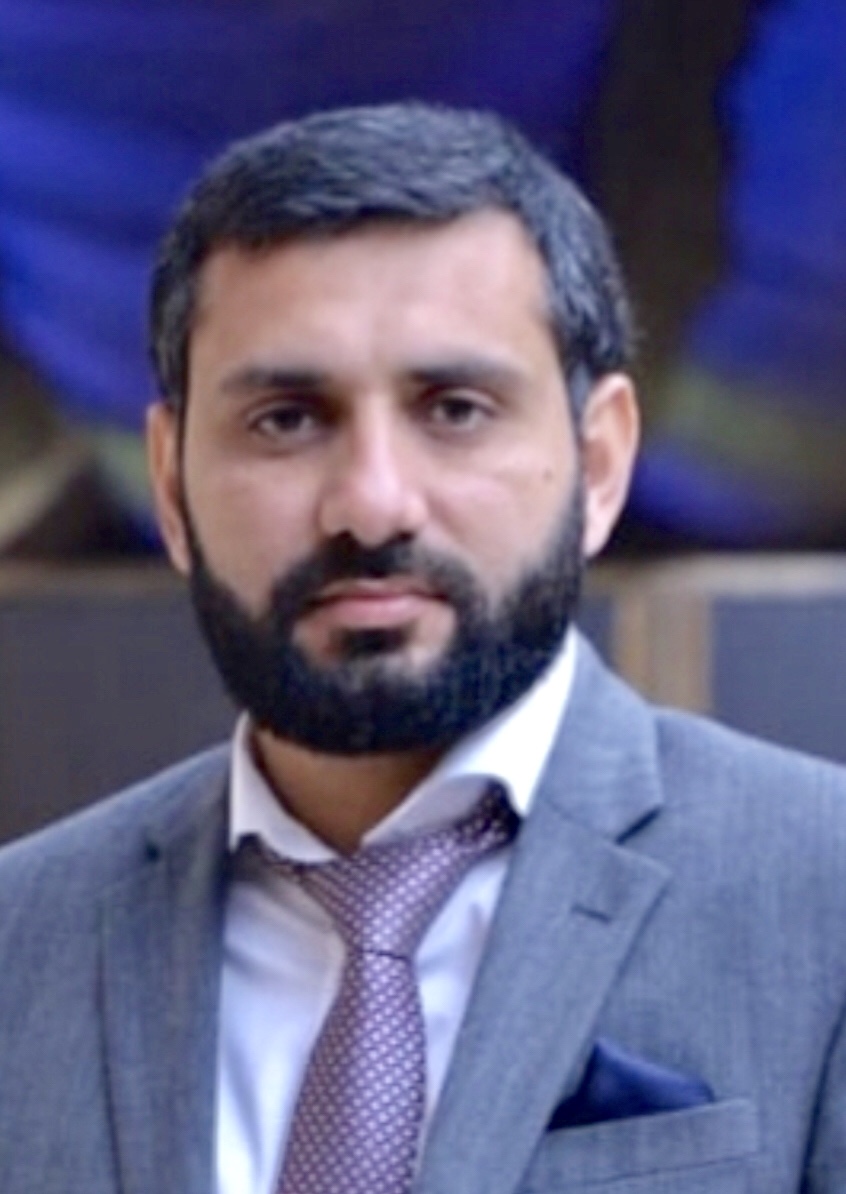}}]{Nasir Saeed} received the bachelor's degree in telecommunication from the University of Engineering and Technology, Peshawar, Pakistan, in 2009, the master's degree in satellite navigation from the Polito di Torino, Italy, in 2012, and the Ph.D. degree in electronics and communication engineering from Hanyang University, Seoul, South Korea, in 2015. He was an Assistant Professor with the Department of Electrical Engineering, Gandhara Institute of Science and IT, Peshawar, from August 2015 to September 2016. He has worked as an Assistant Professor with IQRA National University, Peshawar, from October 2016 to July 2017. From July 2017 to December 2020, he was a Postdoctoral Research Fellow with the Communication Theory Laboratory, King Abdullah University of Science and Technology (KAUST). He is currently an Associate Professor with the Department of Electrical Engineering, National University of Technology (NUTECH), Islamabad, Pakistan. His current research interests include cognitive radio networks, underwater wireless communications, aerial networks, dimensionality reduction, and localization.
\end{IEEEbiography}

\end{document}